\documentclass[10pt,twocolumn,letterpaper]{article}

\usepackage{cvpr}
\usepackage{xcolor}
\usepackage{times}
\usepackage{epsfig}
\usepackage{graphicx}
\usepackage{amsmath}
\usepackage{amssymb}
\usepackage{booktabs}
\usepackage{multirow}
\usepackage{url}
\usepackage{orcidlink}

\begin{document}

\title{Selfie-Capture Dynamics as an Auxiliary Signal Against \\ Deepfakes and Injection Attacks for Mobile Identity Verification}

\author{
Erkka Rantahalvari$^{1}$, Olli Silv\'en$^{1,2,\orcidlink{0000-0002-2661-804X}}$, Zinelabidine Boulkenafet$^{1}$, Constantino \'Alvarez Casado$^{1,2,\orcidlink{0000-0002-3052-4759}}$ \\
\textit{$^{1}$Candour Oy, Oulu, Finland} \\
\textit{$^{2}$University of Oulu, Oulu, Finland}
}

\maketitle

\begin{abstract}
\textcolor{black}{Mobile remote identity verification (RIdV) systems are exposed to attacks that manipulate or replace the facial video stream, including presentation attacks, real-time deepfakes, and video injection. Recent European requirements, including ETSI TS 119 461 and CEN/TS 18099, motivate evidence channels beyond camera-based presentation-attack detection. This paper studies whether passive motion traces recorded during selfie capture provide auxiliary evidence for spoof screening and user verification. We introduce CanSelfie, a dataset of 375 bona fide multi-sensor sequences collected at 50\,Hz from 30 participants using a commercial mobile RIdV application, together with stationary, handheld, and temporally shifted attack-proxy scenarios. We benchmark 7 multivariate time-series classifiers and 8 whole-series anomaly detectors across sensor configurations and temporal windows. For spoof screening, accelerometer-only ROCKAD obtains 0.00\% FRR and 43.8\% FAR, while QUANT+3-NN obtains the lowest FAR of 32.0\% at 2.37\% FRR; both reject all stationary proxies. For same-device and same-session user verification, WEASEL+MUSE reaches 1.07\% EER using 9 channels. Raw acceleration is the most informative modality, and closed-set accuracy alone does not imply good verification because threshold calibration depends on score distributions.}\textcolor{black}{ The findings support selfie-capture motion as a low-friction auxiliary signal, while leaving cross-device, cross-session, and real injection-attack evaluation as necessary next steps. Code and data available at: \url{https://github.com/Ergzar/Selfie-Motion-AD-TSC}.}
\end{abstract}

\vspace{-7mm}
\section{Introduction}
\begin{figure}[ht!]
\begin{center}
\includegraphics[width=1.0\linewidth]{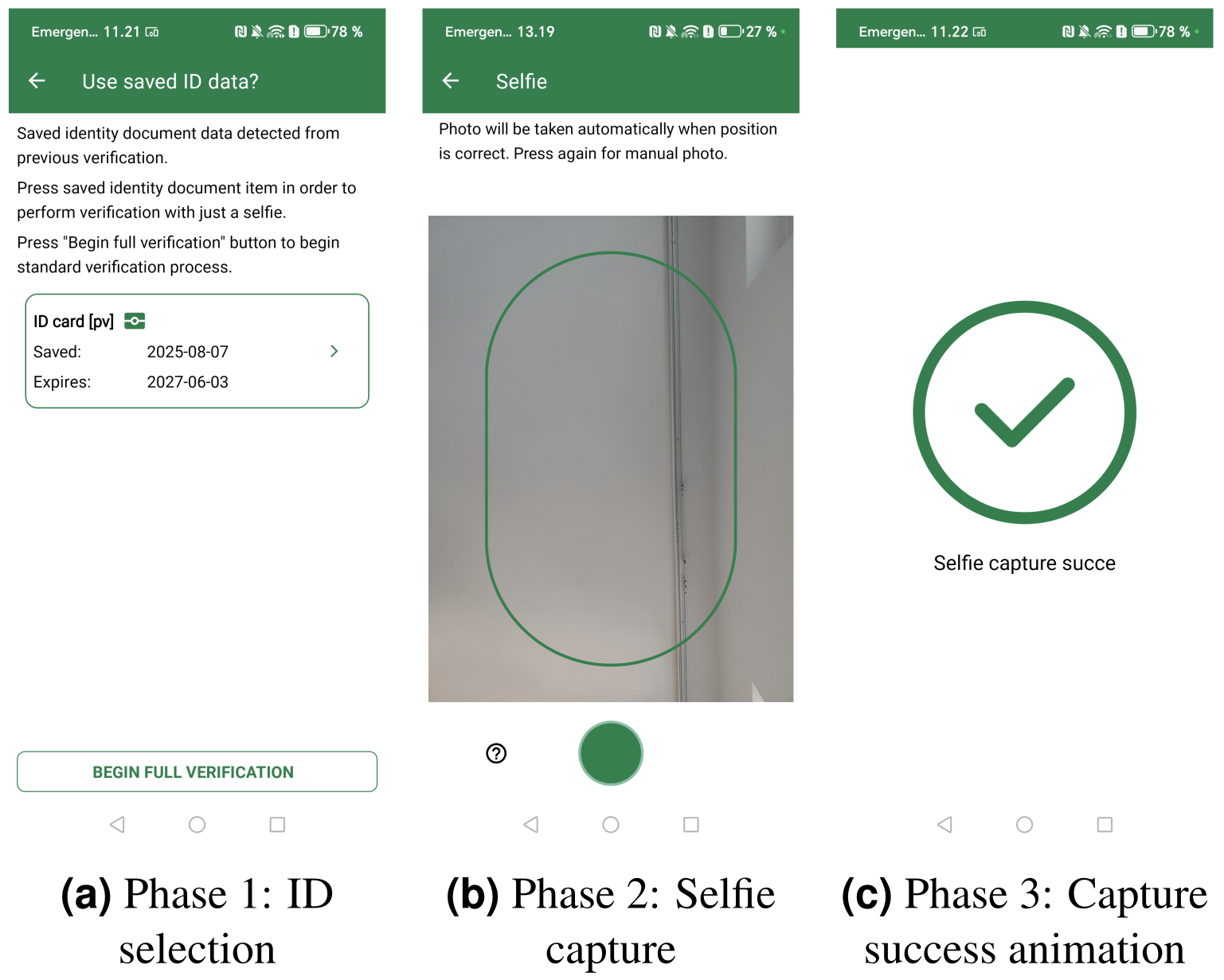}
\end{center}
\vspace{-4mm}
\caption{Three phases of the mobile selfie-capture workflow: (a) ID selection, (b) selfie capture with face alignment, and (c) capture-success animation.}
\label{fig:workflow}
\vspace{-1mm}
\end{figure}

\vspace{-1mm}
Remote identity verification (RIdV) has become a standard requirement for sensitive online services such as banking, government benefits, and border control~\cite{fatima2024large,hahn2026handbook,nanda2023toward}. A typical mobile RIdV workflow captures an identity document, reads its contactless chip when available, and acquires a selfie portrait for comparison with the document portrait, as shown in Figure \ref{fig:workflow}. In workflows involving ICAO-compliant electronic identity documents, the chip can provide a machine-verifiable facial image and signed document data whose authenticity and integrity are verified through the issuer public key infrastructure.

\vspace{-3mm}

The increasing maturity of generative AI has introduced new attack vectors \cite{pei2024deepfake,schmitt2024digital}. Li \etal\ showed that a single photograph can be sufficient to generate an animated face model capable of bypassing representative commercial liveness systems~\cite{li2022seeing}, a finding later echoed by independent red-team reports on commercial KYC providers~\cite{vincent2022deepfake}. When such attacks are combined with rooted or instrumented devices that inject manipulated video in place of the live camera feed~\cite{carta2022video}, the attacker no longer needs to defeat presentation-level defenses, because the synthetic video enters the pipeline as if it were genuine. Mohamed \etal\ also showed that fake motion and position data can be fed to Android applications without rooting the device~\cite{mohamed2017smashed}, although some of those pathways have since been restricted. Open-source tools such as DeepFaceLive~\cite{deepfacelive2024} make real-time face morphing operationally realistic.

These developments are reflected in regulation and standardization. ETSI TS 119 461 for identity proofing of trust-service subjects~\cite{etsi119461}, updated to v2.1.1 in the context of eIDAS~2.0 and the EUDI Wallet~\cite{sharif2024protecting}, introduces Baseline and Extended Levels of Identity Proofing and requires independent testing of biometric injection attack detection according to CEN/TS 18099~\cite{cents18099}. CEN/TS 18099 complements ISO/IEC 30107-3 by defining injection-attack methods, instruments, and an evaluation framework that includes bona fide testing. Together, these documents motivate defense-in-depth in high-assurance RIdV and support complementary signals beyond the facial video stream. One candidate is data from sensing modalities already available on the device. Modern mobile devices contain accelerometers, gyroscopes, and magnetometers~\cite{leimhofer2025crossplatform}, whose signals can be recorded passively during verification. Motion cues do not require additional user action, while settings such as virtual-camera injection in a desktop emulator, replay to a mounted phone, or screen-based presentation may produce traces that are absent, unusually stationary, or temporally inconsistent with handheld selfie capture. Replacing the video feed alone does not reproduce the sensor dynamics of handheld capture. \textcolor{black}{Motion is nevertheless treated only as auxiliary evidence: a real-device attacker may preserve genuine motion traces or jointly manipulate camera and sensor streams.}

\textcolor{black}{This paper investigates whether passive motion traces recorded during selfie capture can support spoof screening and user verification in mobile RIdV. We formulate selfie-capture motion analysis as a biometric security task and introduce CanSelfie, comprising 375 multi-sensor sequences from 30 participants in a commercial RIdV workflow.} \textcolor{black}{The controlled attack-proxy scenarios support a reproducible initial benchmark but do not constitute a complete CEN/TS 18099 injection-attack evaluation.} \textcolor{black}{The main contributions are:}


\begin{itemize}
\item \textcolor{black}{CanSelfie, a publicly available multi-sensor dataset from a commercial selfie-capture workflow, including stationary, handheld, and temporally shifted attack proxies.}

\item \textcolor{black}{A three-task benchmark covering genuine-versus-spoof screening, one-class user verification, and classification-based user verification, with 7 time series classification (TSC) and eight whole-series anomaly detection (AD) methods evaluated across sensor and temporal-window configurations.}

\item \textcolor{black}{An empirical analysis of spoof-related and user-related motion cues, sensor importance, temporal context, and score calibration, identifying the conditions required for cross-device, cross-session, and real injection-attack validation.}
\end{itemize}


\section{Related Work}

This section positions the study at the intersection of remote identity verification, mobile behavioral sensing, and time-series learning. Prior work has studied deepfake and injection threats, mobile sensor-based authentication, and time-series classification and anomaly detection. However, motion-sensor analysis during the selfie-capture phase of mobile RIdV remains scarcely studied as a benchmarked biometric security task.

\subsection{Remote Identity Verification and Spoofing Attacks}

RIdV systems authenticate an identity by comparing a live selfie against the portrait associated with an identity document, often within a workflow that includes document capture and chip-based verification. As discussed by Nanda \etal~\cite{Nanda2024Toward}, modern RIdV is an assurance process combining document evidence, biometric comparison, device-side capture, and workflow integrity.

Recent work shows this threat model has expanded with generative AI and software-level manipulation~\cite{pei2024deepfake,schmitt2024digital}. Li \etal~\cite{li2022seeing} showed that a single photograph may generate an animated face capable of bypassing representative liveness systems, while Carta \etal~\cite{carta2022video} demonstrated that a rooted and instrumented Android device can inject pre-recorded or synthetic video in place of the live camera feed of an RIdV application. Mohamed \etal~\cite{mohamed2017smashed} showed that Android sensor streams can be manipulated at the application level without rooting the device. Reviews of smartphone face authentication show that attacks include printed photos, replayed videos, 3D masks, deepfakes, adversarial software manipulation, and capture-pipeline compromise~\cite{Zheng2023Spoofing,yu2023review,pei2024deepfake}. This threat landscape has motivated defenses beyond RGB selfie analysis. Some methods remain camera-based, such as FaceCloseup~\cite{Li2025FaceCloseup}, which uses perspective distortion cues from close-range facial video. Other studies use complementary channels, such as SonarGuard~\cite{Zhang2023SonarGuard}, which applies ultrasonic sensing to distinguish live faces from spoofing artifacts. At the system level, Kira \etal~\cite{Kira2025Trustworthy} argued for multi-layered trustworthy face recognition services that combine presentation-attack mitigation, integrity checks, and broader system protections. CEN/TS 18099~\cite{cents18099} and ETSI TS 119 461~\cite{etsi119461} follow the same direction by emphasizing injection attack detection and evidence beyond the facial image stream. Yet little work has studied whether motion traces recorded during selfie capture can support spoof screening in mobile RIdV.

\subsection{Motion Sensors and Behavioral Biometrics on Mobile Devices}

Smartphones typically include a triaxial accelerometer, gyroscope, and, in many cases, magnetometer~\cite{leimhofer2025crossplatform}. These signals form multivariate sequences that reflect device handling, orientation changes, and user interaction patterns~\cite{lima2019human,zhou2020how}. Sampling rates around 50\,Hz are generally sufficient for human hand activity in mobile settings~\cite{alvarez2021meditation,sitova2015hmog,yamane2025effects}.

Within behavioral biometrics, mobile signals have mainly been studied for continuous or repeated authentication. Mahbub \etal~\cite{mahbub2019continuous} modeled application-usage sequences with Markovian and hidden Markov models, showing that temporal smartphone behavior can provide identity evidence despite challenges such as sparse sessions and unseen observations. Sitov\'{a} \etal~\cite{sitova2015hmog} introduced HMOG, where accelerometer and gyroscope signals during typing and interaction were used to derive grasp resistance and grasp stability. Estrela \etal~\cite{estrela2021framework} studied touch-dynamics authentication for mobile banking, while Qiao \etal~\cite{qiao2025contrastive} and Shen \etal~\cite{shen2022mmauth} studied multimodal behavioral fusion. These works confirm that mobile interaction patterns and sensor signals carry identity-related information, but their setting differs from RIdV selfie capture, which is short, structured, task-specific, and directly relevant to spoof screening.

\subsection{Time Series Classification}
\label{sec:related_tsc}

\textcolor{black}{Time series classification has progressed substantially during the last decade~\cite{alvarez2021meditation,bagnall2017great}. Bagnall \etal~\cite{bagnall2017great} established strong baselines across many datasets and showed the advantage of ensemble methods over 1-NN DTW. More recently, Middlehurst \etal~\cite{middlehurst2024bakeoffredux} confirmed the competitiveness of transform-based and convolution-based methods, including the Hydra family~\cite{dempster2023hydra}. Many effective TSC methods follow a transform-then-classify design, where the raw sequence is mapped into a discriminative feature space before classification~\cite{middlehurst2024bakeoffredux}.}

\textcolor{black}{Relevant families for short multivariate motion traces include ROCKET-like random-kernel methods~\cite{dempster2020rocket}, dictionary methods such as WEASEL+MUSE for multivariate symbolic words~\cite{schafer2018weaselmuse}, interval methods such as QUANT~\cite{dempster2024quant} and r-STSF~\cite{cabello2024fast}, shapelet methods such as RDST~\cite{guillaume2022random}, compact feature methods such as catch22~\cite{lubba2019catch22}, and deep models such as ResNet~\cite{wang2017time}. This range is relevant because selfie-capture motion is short, multivariate, and structured by both user-specific handling and task-specific capture dynamics.}

\subsection{Time Series Anomaly Detection}

Time-series anomaly detection has traditionally focused on point anomalies or anomalous subsequences~\cite{schmidl2022anomaly,chandola2009anomaly}. Here, the decision unit is the full motion sequence associated with one selfie-capture attempt, which places the task closer to whole-series anomaly detection. ROCKAD~\cite{theissler2023rockad} is relevant because it combines ROCKET-based features, power transformation, and a distance-based detector. Classical one-class methods such as Isolation Forest~\cite{liu2008isolation} and one-class SVM~\cite{scholkopf2001estimating,tax2004support} remain important because they can be trained only on bona fide samples, and Euclidean or DTW distance-based methods are natural when few enrollment samples are available. In biometric verification, such one-class formulations measure how strongly a probe deviates from the distribution or neighborhood structure of enrolled genuine samples~\cite{jain2011introduction}, making whole-series anomaly detection a suitable complement to closed-set verification.

\vspace{-2mm}
\section{Methodology}
\label{sec:methodology}

\vspace{-1mm}

\subsection{Problem Setting and Threat Model}

\textcolor{black}{We consider a mobile RIdV workflow where a user captures an identity document and a selfie image or short selfie sequence for comparison with the document portrait. The security problem concerns the selfie-capture phase, where an attacker may present a spoof, replay a face from another display, or replace the camera stream through software-level injection. In such cases, the facial video stream may appear plausible while device motion is absent, unusually stationary, or temporally inconsistent with handheld capture. The motion layer is treated as auxiliary evidence: it does not replace face recognition or presentation-attack detection, and attacks that jointly forge camera and motion streams are outside the benchmark scope. }\textcolor{black}{In particular, a strong software-level injection attack on a rooted or instrumented device can replace the camera feed while preserving authentic device motion, so such attacks cannot be rejected by motion evidence alone and are explicitly out of scope. }\textcolor{black}{The objective is to test whether passive motion traces from a real selfie-capture workflow contain measurable information for spoof screening and auxiliary user verification.}


\subsection{System Overview}
\begin{figure}[ht!]
\begin{center}
\includegraphics[width=0.99\linewidth]{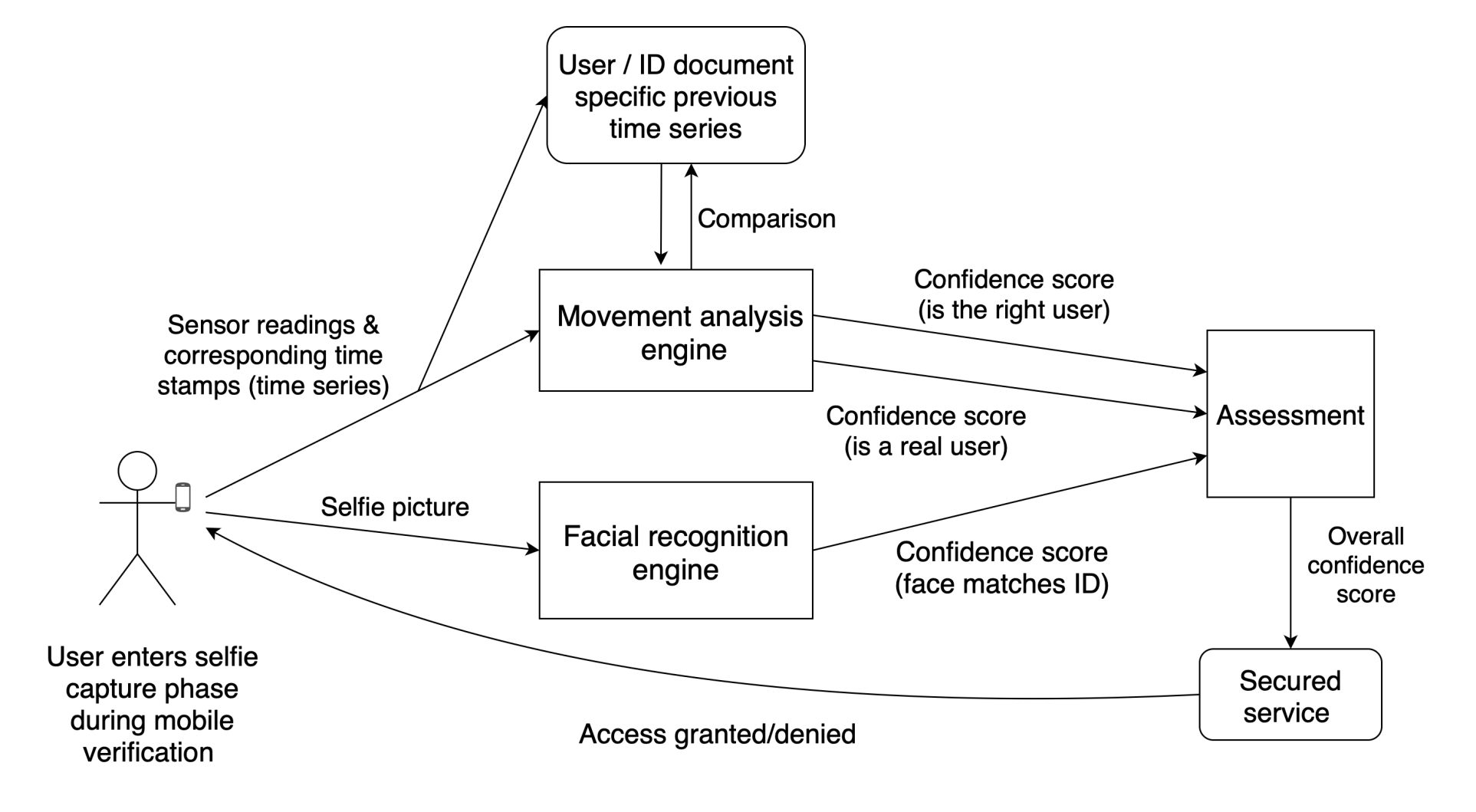}
\end{center}
\vspace{-5mm}
\caption{Motion analysis as an auxiliary component in the identity verification pipeline. Sensor readings collected during selfie capture are converted into a confidence score that supports the final decision. \textcolor{black}{Only the motion branch is evaluated; fusion is outside the present experiments.}}
\vspace{-4mm}
\label{fig:system}
\end{figure}

\vspace{-1mm}
\textcolor{black}{Figure~\ref{fig:system} shows the intended role of motion analysis in the RIdV pipeline. During selfie capture, sensor readings are passed to a motion analysis engine, which outputs a confidence score indicating whether the observed motion is consistent with bona fide capture behavior or with the claimed user, depending on the task. The decision logic is asymmetric: a facial recognition rejection is not overridden by motion, but a low motion confidence score can increase the evidence for rejection or step-up verification. This design prioritizes low false rejection for bona fide users while allowing the motion channel to reject part of the attack space. This role is consistent with the defense-in-depth direction of ETSI TS 119 461~\cite{etsi119461} and CEN/TS 18099~\cite{cents18099}, while this study remains an empirical benchmark rather than a formal conformance evaluation. }\textcolor{black}{No video-motion fusion or end-to-end face/motion decision fusion is evaluated in this paper.}

\subsection{CanSelfie Dataset}

\textcolor{black}{Mobile face presentation-attack datasets such as SOTERIA~\cite{ramoly2024soteria} pair face videos with inertial motion and, on compatible devices, depth data, but they are not centered on the selfie-capture event of a commercial RIdV workflow and do not provide a stationary, handheld, and temporal-shift attack-proxy protocol for passive motion.}  \textcolor{black}{No public dataset matched the specific passive selfie-capture motion benchmark considered here. We therefore collected CanSelfie, a multi-sensor dataset recorded with a modified commercial RIdV application. The application recorded motion data in the background during selfie capture while preserving normal user interaction flow. The sensor polling rate was 50\,Hz.}

Five three-axis sensor streams were recorded: raw acceleration, gyroscope angular velocity, magnetometer field strength, linear acceleration with gravity removed, and the gravity vector, giving 15 channels. Unless otherwise stated, the main benchmark uses the 9 physical channels corresponding to raw acceleration, gyroscope, and processed magnetometer. Linear acceleration and gravity are kept for sensor-ablation analysis because they are derived from acceleration and device attitude estimates. Data were collected from 30 participants, including 20 male and 10 female participants, using one Huawei Honor 200 smartphone. Each participant performed 10 to 15 selfie-capture sequences in a sitting position within a single session. The workflow, shown in Figure~\ref{fig:workflow}, consisted of selecting the identity document option, opening the selfie camera, aligning the face, waiting for auto-capture or pressing the shutter button, observing the capture-success animation, and waiting for the server result. The public release contains motion traces and benchmark metadata, but no facial images, identity-document images, names, or direct identifiers. Participants provided their informed consent for the collection and use of data for research purposes.

\subsection{Attack-Proxy Scenarios}

\begin{figure}[ht!]
\vspace{-3mm}
\begin{center}
\includegraphics[width=0.99\linewidth]{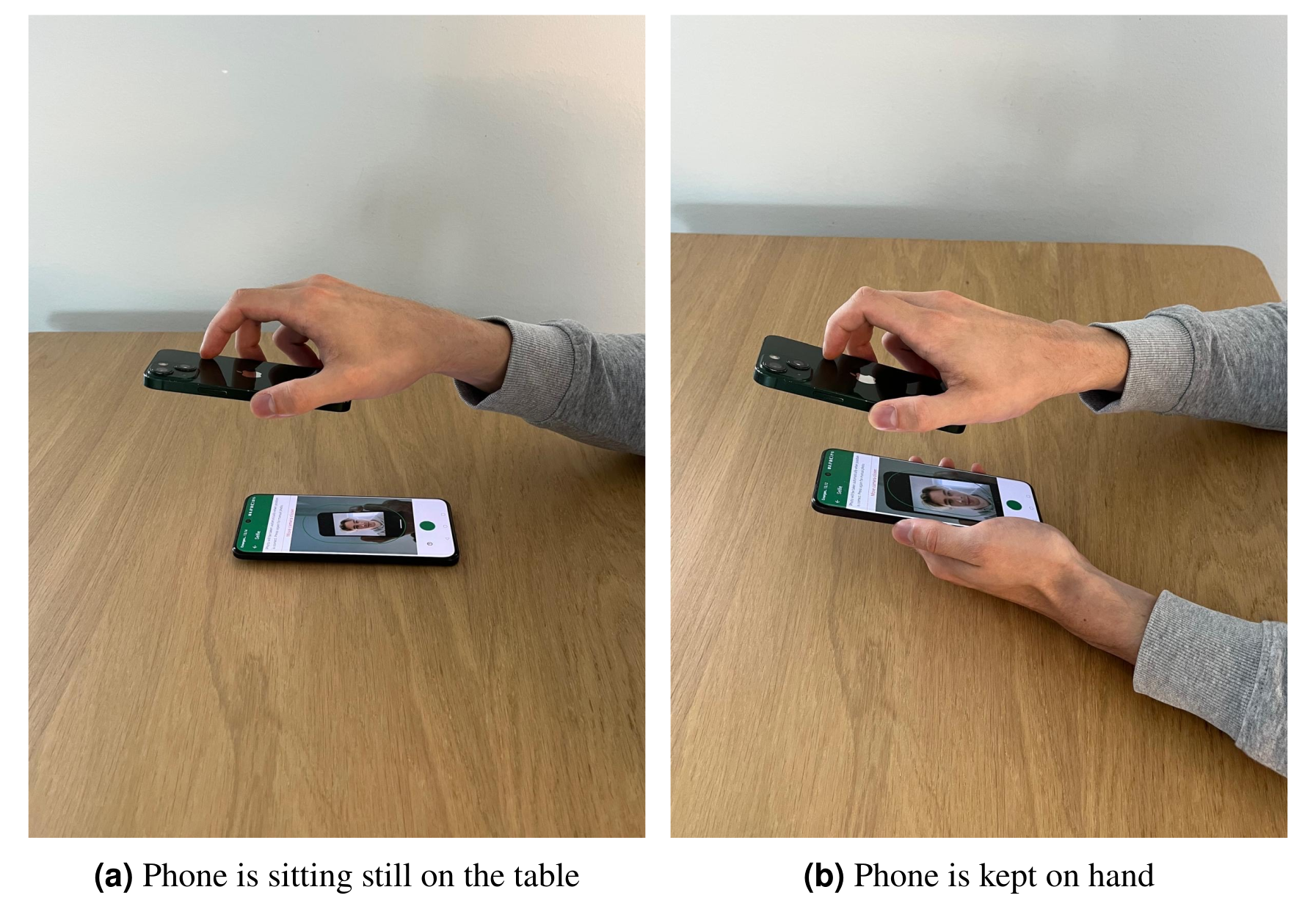}
\end{center}
\vspace{-4mm}
\caption{Attack-proxy scenarios: (a) stationary replay with the phone on a table, and (b) handheld replay with another screen presented to the camera.}
\label{fig:spoofs}
\vspace{-3mm}
\end{figure}

We define three attack-proxy scenarios for spoof screening. They do not cover the full space of presentation or software-level injection attacks, but approximate cases where camera evidence and device motion may become inconsistent. In the stationary replay proxy, the phone was placed on a table or mount while a face photograph was displayed to the camera from another device, producing facial content without normal handheld motion. Six sequences were recorded. In the handheld replay proxy, the phone was held while a face photograph or video was presented from another screen, as shown in Figure~\ref{fig:spoofs}. Eleven sequences were recorded from two participants. In the temporal-shift proxy, 18 bona fide sequences from three unseen participants were used with the selfie-capture timestamp shifted by 1.5 to 3 seconds, representing temporal inconsistency that may occur in real-time morphing or video-injection settings~\cite{carta2022video,li2022seeing}.

\subsection{Preprocessing and Window Extraction}

Let $X_i \in \mathbb{R}^{T_i \times C}$ denote the raw motion sequence of capture attempt $i$, where $T_i$ is the number of samples and $C$ is the number of sensor channels. Each bona fide sequence has a participant label $y_i \in \{1,\ldots,N\}$, with $N=30$. Attack-proxy sequences are assigned an attack type and are not used for training in the semi-supervised spoof-screening setting.

Magnetometer readings contained a heading-related bias associated with participant direction. To reduce this context leakage while preserving relative heading changes, magnetometer vectors were centered by the sequence-level mean, normalized to unit length, and converted into first-order temporal differences. A second-order 12.5\,Hz low-pass Butterworth filter was applied to reduce high-frequency noise~\cite{sa2022filters}. Uniform-length samples were extracted using the selfie-capture timestamp and camera-opening timestamp. The selfie-centered window $W_c$ contains 50 samples before and 150 samples after capture, giving 200 samples, or approximately 4 seconds at 50\,Hz. The camera-opening window $W_o$ contains the first $K=10$ samples after the selfie camera opens. The three temporal representations are $W_c$ alone, $W_o$ concatenated with $W_c$ along time, and a double-window representation in which $W_o$ and $W_c$ are stacked as parallel channels.

\subsection{Benchmark Tasks}
\vspace{-1mm}
The benchmark contains three tasks. In genuine-versus-spoof screening, models are trained only on bona fide sequences and evaluated on held-out bona fide sequences and attack proxies. In 10-shot one-class user verification, each enrolled user is represented by 10 bona fide sequences, and impostor probes are zero-effort samples from other participants. In classification-based user verification, a closed-set TSC identifies participants, and the predicted probability for the claimed user is used as the verification score.

\section{Experimental Evaluation}

\subsection{Benchmarked Methods}

\textbf{Time series classification.} Based on Section~\ref{sec:related_tsc}, seven TSC algorithms were selected following the 2024 bake-off taxonomy~\cite{middlehurst2024bakeoffredux}: MR-HYDRA~\cite{dempster2023hydra}, WEASEL+MUSE~\cite{schafer2018weaselmuse}, QUANT~\cite{dempster2024quant}, r-STSF~\cite{cabello2024fast}, RDST~\cite{guillaume2022random}, ResNet~\cite{wang2017time}, and catch22~\cite{lubba2019catch22}. The selection covers hybrid convolution-dictionary, dictionary-based, interval-based, shapelet-based, deep learning, and feature-based designs. Implementations were obtained from \textit{aeon}~\cite{middlehurst2024aeon}.

\textbf{Whole-series anomaly detection.} Eight anomaly detectors were evaluated: ROCKAD~\cite{theissler2023rockad}, Isolation Forest on raw series~\cite{liu2008isolation}, Isolation Forest on QUANT features, one-class SVM with an RBF kernel~\cite{scholkopf2001estimating,tax2004support}, Euclidean $k$-NN, DTW $k$-NN, QUANT-based $k$-NN, and an LSTM autoencoder~\cite{malhotra2016lstm}. These represent complementary full-sequence anomaly scores: distances to bona fide references, support estimation, tree-based isolation, feature-space detection, and reconstruction error. Table~\ref{tab:detectors} summarizes the detectors.

\subsection{Sensor and Window Configurations}
\vspace{-1mm}
Four sensor configurations were evaluated: accelerometer $x$-axis only, all accelerometer axes, the $x$-axis of accelerometer, gyroscope, and processed magnetometer, and all 9 axes from these three sensors. Additional ablations with linear acceleration, gravity, and processed magnetometer analyze derived and physical sensor streams. The windowing experiments compare the three representations in Section~\ref{sec:methodology}: $W_c$ alone, temporal concatenation of $W_o$ and $W_c$, and the double-window representation.

\begin{table}[ht!]
\def\arraystretch{0.94}
\begin{center}
\caption{Anomaly detection methods, categories, and implementation origins.}
\vspace{-2mm}
\label{tab:detectors}
\small
\begin{tabular}{lll}
\toprule
Method & Category & Source \\
\midrule
ROCKAD & Distance (ROCKET) & \cite{theissler2023rockad} \\
OCSVM (RBF) & Support vector & scikit-learn \\
Isolation Forest & Tree ensemble & scikit-learn \\
Euclidean $k$-NN & Distance & Custom \\
$k$-NN DTW & Distance & aeon \\
QUANT + $k$-NN & Distance & aeon \\
QUANT + IF & Tree ensemble & aeon \\
LSTM-AE & Autoencoder & \textcolor{black}{Custom (PyTorch)} \\
\bottomrule
\end{tabular}
\vspace{-7mm}
\end{center}

\end{table}

\subsection{Training, Calibration, and Testing Protocols}

All thresholds and fitted transformations were estimated only from training or calibration data. Held-out test subjects and attack-proxy samples were not used for threshold selection.

\begin{figure}[ht!]
\begin{center}
\includegraphics[width=0.99\linewidth]{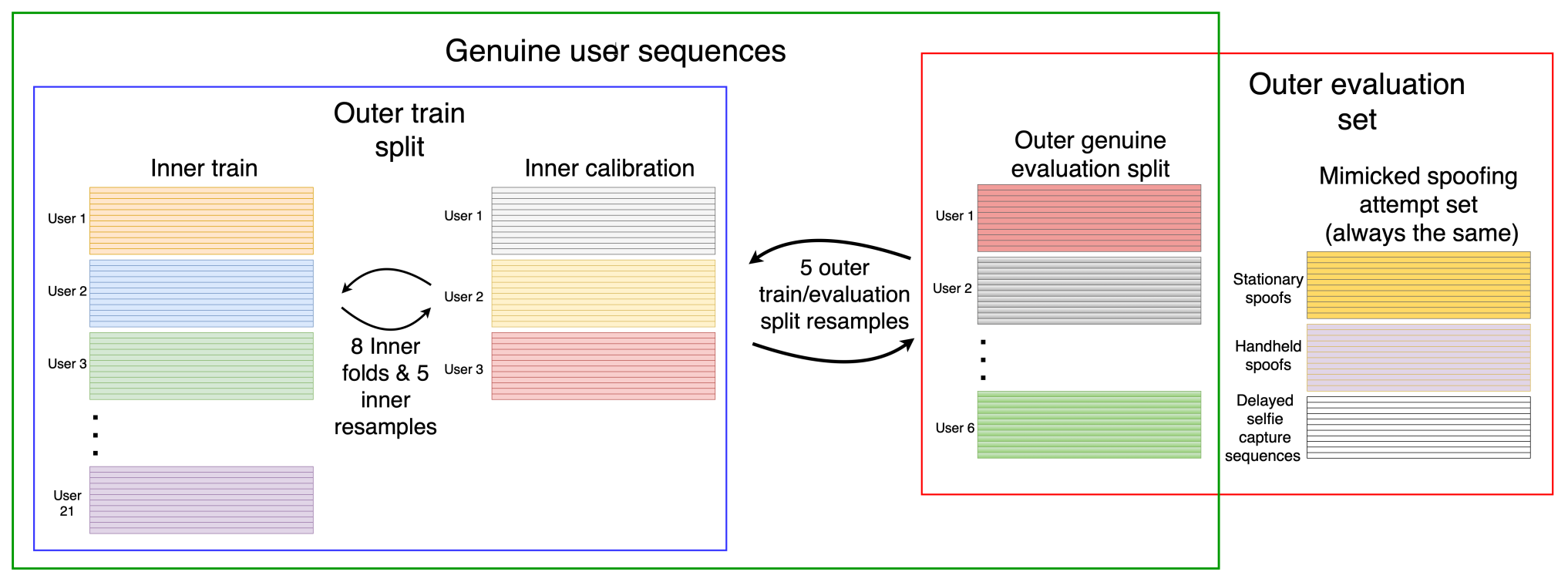}
\end{center}
\vspace{-4mm}
\caption{Partitioning protocol for genuine-versus-spoof screening. Bona fide sequences are split at participant level; inner folds calibrate the threshold; attack-proxy sequences are used only for evaluation.}
\label{fig:protocol}
\end{figure}

\textbf{Genuine-versus-spoof screening.} Anomaly detectors were trained only on bona fide sequences. Group-aware splitting ensured that all sequences from one participant appeared either in training or in the bona fide test partition. Within training, out-of-fold bona fide scores set a global threshold $\tau$ at the 99th percentile of the anomaly-score distribution, targeting FRR $\approx 1\%$. The threshold was applied to held-out bona fide samples and to the fixed attack-proxy set. The procedure was repeated five times with participant-level splits, as shown in Figure~\ref{fig:protocol}.

\vspace{-2mm}
\vspace{-2mm}

\textbf{10-shot one-class user verification.} For each participant, 10 bona fide sequences were used for enrollment and the remaining 2 to 5 sequences as genuine probes. Impostor probes were bona fide sequences from other participants. Detectors were trained independently per enrolled user, and a user-specific threshold $\tau_u$ was calibrated from 2-fold inner cross-validation repeated five times.

\textbf{Classification-based user verification.} Time series classifiers were trained as closed-set identification models with 5-fold stratified outer cross-validation. Within each training fold, out-of-fold genuine scores were computed with up to 3 inner folds and 5 repeats to calibrate a global threshold $\tau$ targeting FRR = 1\%. A verification trial for claimed user $u$ was scored using the predicted probability assigned to $u$. EER was computed by threshold sweeping within each outer fold and macro-averaging across users.

\subsection{Evaluation Metrics}

The primary metrics are false rejection rate (FRR), false acceptance rate (FAR), and equal error rate (EER), following ISO/IEC 19795-1:2021~\cite{isoiec19795_1_2021}. FRR measures rejected bona fide or genuine trials, FAR measures accepted attack-proxy or impostor trials, and EER is obtained by sweeping the decision threshold and finding the point where both rates are equal. For spoof screening, FRR and FAR correspond conceptually to Bona fide Presentation Classification Error Rate (BPCER) and Attack Presentation Classification Error Rate (APCER), respectively. Since the attack set consists of controlled proxies rather than a formal ISO/IEC 30107-3 or CEN/TS 18099 test suite, we report FRR and FAR with per-attack-type FAR. For verification experiments, EER is computed by threshold sweeping, whereas FRR and FAR are reported at a training-calibrated threshold targeting FRR = 1\%. The two values can therefore differ substantially: a method with a low EER may still produce a high FAR at the calibrated operating point if its score distribution is poorly graduated near low-FRR thresholds. Overall spoof FAR is computed as the equal-weight mean across stationary, handheld, and temporal-shift proxies.

\subsection{Implementation Details}

Experiments were implemented in Python 3.10 using \textit{aeon} for TSC models~\cite{middlehurst2024aeon}, scikit-learn for data partitioning, classical classifiers, Isolation Forest, OCSVM, nearest-neighbor routines, and evaluation utilities, PyTorch for the LSTM autoencoder, and SciPy for preprocessing. Unless otherwise stated, classifier hyperparameters followed the default \textit{aeon} configurations; ResNet was trained for 150 epochs with batch size 32; ROCKAD used 24 estimators, 1024 random kernels, 3 neighbors, and power transformation; Isolation Forest used 1500 trees with automatic sample size and contamination; OCSVM used $\nu=0.05$ and $\gamma=\texttt{scale}$; the LSTM autoencoder used latent dimension 24, batch size 32, 50 epochs, validation split 0.20, early-stopping patience 6, and $L_2=5 \times 10^{-4}$; QUANT detectors used interval depth 6 and quantile divisor 4; and nearest-neighbor detectors used $k=3$. Experiments used random seed 7 and ran on a workstation with an AMD\textsuperscript{\textregistered} Ryzen 3700X 8-core processor, 64\,GB RAM, and two NVIDIA GeForce\textsuperscript{\textregistered} RTX 2080 GPUs. Results are reported as mean and standard deviation.

\section{Results}

\subsection{Motion Patterns During Selfie Capture}

\begin{figure}[ht!]
\begin{center}
\includegraphics[width=0.99\linewidth]{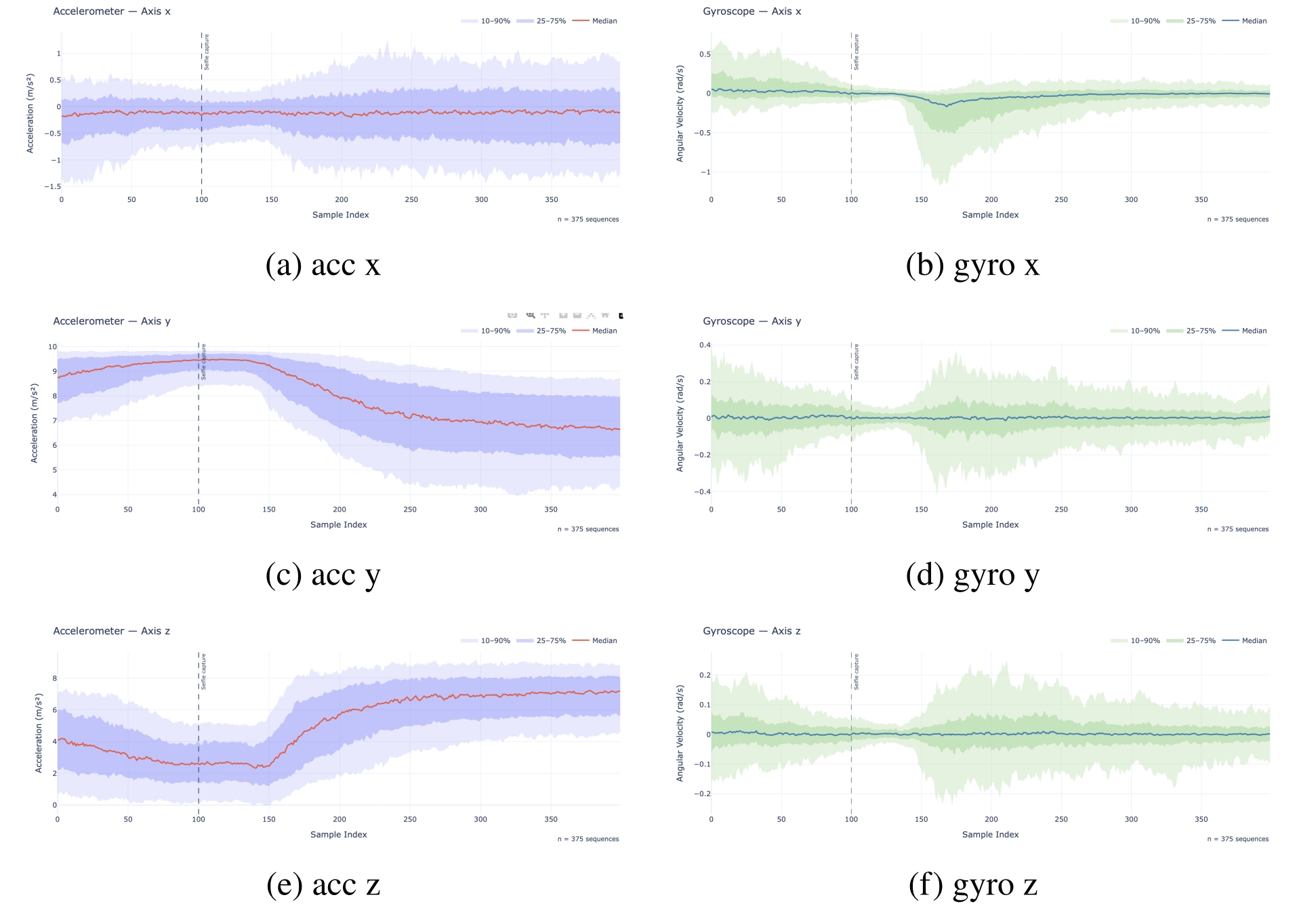}
\end{center}
\vspace{-3mm}
\caption{Percentile bands around the selfie-capture moment for accelerometer and gyroscope axes. Darker shading shows the 25th--75th percentiles and lighter shading the 10th--90th percentiles across 375 bona fide sequences.}
\label{fig:percentile}
\end{figure}

Figure~\ref{fig:percentile} shows median time series with percentile bands for accelerometer and gyroscope axes aligned to the selfie-capture moment. Before capture, users tend to move the phone toward a more vertical position, visible as a decrease in the accelerometer $z$-axis component. After capture, the device remains relatively stable for about one second, likely during the success animation, and is then tilted back toward a more horizontal position, with a corresponding rotation in the gyroscope $x$-axis. Motion variability increases after one to two seconds as users wait for the server response. These patterns show that selfie capture contains structured temporal motion rather than arbitrary handheld movement.

\vspace{-2mm}
\vspace{-3mm}

\subsection{Genuine-versus-Spoof Screening}

Table~\ref{tab:spoof_acc} reports spoof-screening results for the 3-axis accelerometer configuration. ROCKAD achieved 0.00\% FRR with 43.8\% overall FAR, rejecting a substantial fraction of attack proxies without rejecting bona fide samples in this evaluation. QUANT+3-NN obtained the lowest overall FAR, 32.0\%, at 2.37\% FRR. Isolation Forest on raw series also obtained low FAR, 35.6\%, but with higher FRR variability.

\begin{table}[ht!]
\def\arraystretch{0.97}
\setlength{\tabcolsep}{0.9em}
\begin{center}
\caption{Spoof-screening results for the 3-axis accelerometer configuration using a 10+50+100 window. Overall FAR is the equal-weight mean over the three attack-proxy types. Values are mean $\pm$ standard deviation over 5 resamples.}
\label{tab:spoof_acc}
\small
\vspace{-3mm}
\begin{tabular}{lccc}
\toprule
Method & FRR (\%) & FAR (\%) & Delay (ms) \\
\midrule
ROCKAD        & 0.00$\pm$0.00 & 43.8$\pm$0.8 & 3.48 \\
QUANT 3-NN    & 2.37$\pm$1.46 & 32.0$\pm$0.0 & 0.46 \\
QUANT-IF      & 0.00$\pm$0.00 & 91.4$\pm$0.0 & 0.52 \\
IF (raw)      & 4.81$\pm$8.04 & 35.6$\pm$2.5 & 0.13 \\
OCSVM         & 0.81$\pm$1.81 & 43.2$\pm$3.1 & 0.00 \\
Eucl. 3-NN    & 0.81$\pm$1.81 & 72.2$\pm$0.0 & 0.03 \\
LSTM-AE       & 1.33$\pm$1.66 & 41.4$\pm$0.8 & 0.27 \\
\bottomrule
\end{tabular}
\end{center}
\vspace{-3mm}
\end{table}

Table~\ref{tab:spoof_type} shows that performance depends strongly on attack-proxy type. Stationary replay proxies were detected by most methods with FAR = 0.0\%. Handheld replay proxies were more difficult, although QUANT+3-NN reduced FAR to 18.2\%. Temporal-shift proxies were the hardest condition, with the best FAR remaining 67.8\%, because they contain genuine handheld motion and differ mainly through the shifted capture timestamp.

\begin{table}[ht!]
\def\arraystretch{1.0}
\setlength{\tabcolsep}{1.2em}
\begin{center}
\caption{FAR (\%) by attack-proxy type for the 3-axis accelerometer configuration. Stationary: phone on table. Handheld: phone held while showing another screen. Delayed: bona fide sequences with shifted capture timestamp.}
\label{tab:spoof_type}
\small
\begin{tabular}{lccc}
\toprule
Method & Static & Handheld & Delayed \\
\midrule
ROCKAD       & 0.0 & 63.6 & 67.8 \\
QUANT 3-NN   & 0.0 & 18.2 & 77.8 \\
IF (raw)     & 0.0 & 29.1 & 77.8 \\
OCSVM        & 0.0 & 47.3 & 82.2 \\
LSTM-AE      & 0.0 & 36.4 & 87.8 \\
\bottomrule
\end{tabular}
\end{center}
\vspace{-5mm}
\end{table}

The 9-channel configuration did not consistently improve spoof screening over the accelerometer-only setting. QUANT+3-NN again obtained the lowest overall FAR, 32.0\%, with 2.38\% FRR, while the 3-channel configuration using only the $x$-axis of each sensor performed poorly. Thus, raw acceleration carries most spoof-screening signal, and adding channels does not automatically improve robustness.

\subsection{10-Shot One-Class User Verification}
\vspace{-1mm}
Table~\ref{tab:oneclass} reports 10-shot one-class user verification using the $x$-axis of the accelerometer, gyroscope, and magnetometer. Euclidean 3-NN achieved 0.00\% FRR with 79.94\% FAR, while DTW 3-NN achieved 0.87\% FRR with 74.77\% FAR. Low-shot one-class models preserve bona fide acceptance but reject only a limited fraction of zero-effort impostors, so this setting provides weak auxiliary user-specific evidence and is not sufficient as standalone verification.

\begin{table}[ht!]
\def\arraystretch{1.0}
\setlength{\tabcolsep}{1.9em}
\begin{center}
\caption{10-shot one-class user verification using $x$-axis accelerometer, gyroscope, and magnetometer with a 10+50+150 window. Thresholds are calibrated per user using 2 inner folds.}
\label{tab:oneclass}
\small
\vspace{-3mm}
\begin{tabular}{lcc}
\toprule
Method & FRR (\%) & FAR (\%) \\
\midrule
OCSVM (RBF)    & 13.04 & 58.64 \\
Euclidean 3-NN & 0.00  & 79.94 \\
3-NN DTW       & 0.87  & 74.77 \\
QUANT + 3-NN   & 3.19  & 79.98 \\
\bottomrule
\end{tabular}
\end{center}
\vspace{-3mm}
\end{table}

\subsection{TSC-Based User Verification}

Table~\ref{tab:tsc_univariate} compares TSC-based verification using the univariate accelerometer $x$-axis. The double-window representation consistently reduced EER compared with the single selfie-centered window. QUANT and WEASEL+MUSE achieved the best EER in this setting, both reaching 8.0\%, showing that initial handling dynamics after camera opening provide complementary information.

Table~\ref{tab:tsc_multi} shows the effect of adding channels under the double-window setting. Moving from one accelerometer axis to the $x$-axis of three sensors improved all methods. Using all accelerometer axes further improved most methods, with WEASEL+MUSE reaching 1.29\% EER. The full 9-channel setting produced the best result, 1.07\% EER with WEASEL+MUSE, although the gain over the 3-axis accelerometer setting was small.

\vspace{-1mm}
\begin{table}[ht!]
\def\arraystretch{0.97}
\setlength{\tabcolsep}{1.1em}
\begin{center}
\caption{TSC-based verification with the accelerometer $x$-axis: single selfie-centered window (S) versus double window (D). Results are computed with 5-fold cross-validation. FAR is reported at the calibrated low-FRR operating point.}
\label{tab:tsc_univariate}
\small
\vspace{-3mm}
\begin{tabular}{lcccc}
\toprule
 & \multicolumn{2}{c}{EER (\%)} & \multicolumn{2}{c}{FAR (\%)} \\
\cmidrule(lr){2-3} \cmidrule(lr){4-5}
Method & S & D & S & D \\
\midrule
QUANT         & 12.6 & 8.0  & 68.1 & 66.6 \\
MR-HYDRA      & 30.6 & 18.6 & 100  & 100 \\
WEASEL+MUSE   & 16.6 & 8.0  & 46.6 & 36.6 \\
r-STSF        & 12.8 & 9.3  & 68.7 & 55.7 \\
ResNet        & 20.1 & 14.4 & 55.9 & 50.7 \\
catch22       & 17.2 & 10.3 & 75.0 & 54.3 \\
\bottomrule
\end{tabular}
\end{center}
\vspace{-5mm}
\end{table}

\begin{table}[ht!]
\def\arraystretch{0.95}
\setlength{\tabcolsep}{1.5em}
\begin{center}
\caption{TSC-based verification with multiple channel configurations and the double-window representation. Configurations: $x$-axes of three sensors (3ch), all accelerometer axes (3acc), and all axes from all sensors (9ch).}
\label{tab:tsc_multi}
\small
\vspace{-3mm}
\begin{tabular}{lccc}
\toprule
 & \multicolumn{3}{c}{EER (\%)} \\
\cmidrule(lr){2-4}
Method & 3ch & 3acc & 9ch \\
\midrule
QUANT         & 6.30 & 5.17 & 5.06 \\
MR-HYDRA      & 15.80 & 9.42 & 9.85 \\
WEASEL+MUSE   & 2.00 & 1.29 & 1.07 \\
r-STSF        & 6.10 & 5.21 & 5.09 \\
ResNet        & 9.81 & 8.46 & 6.76 \\
catch22       & 7.26 & 4.35 & 4.85 \\
\bottomrule
\end{tabular}
\end{center}
\vspace{-3mm}
\end{table}

Table~\ref{tab:tsc_best} reports 9-channel double-window results with repeated 5-fold cross-validation. WEASEL+MUSE achieved the strongest verification performance, with 1.07\% EER, 1.40\% FRR, and 7.01\% FAR. MR-HYDRA produced FAR = 100\% at the calibrated threshold, showing that closed-set classification performance does not necessarily translate into useful verification scores.

\begin{table}[ht!]
\def\arraystretch{0.95}
\setlength{\tabcolsep}{0.7em}
\begin{center}
\caption{TSC-based verification using 9 channels and the double-window representation. EER is computed by threshold sweeping; FRR and FAR are reported at the training-calibrated threshold (target FRR = 1\%). Values are mean $\pm$ standard deviation over repeated 5-fold cross-validation.}
\label{tab:tsc_best}
\small
\begin{tabular}{lccc}
\toprule
Method & EER (\%) & FRR (\%) & FAR (\%) \\
\midrule
QUANT         & 5.06$\pm$0.50 & 1.07$\pm$0.48 & 57.9$\pm$2.8 \\
MR-HYDRA      & 9.85$\pm$0.38 & 0.00$\pm$0.00 & 100$\pm$0.0 \\
WEASEL+MUSE   & 1.07$\pm$0.09 & 1.40$\pm$0.19 & 7.01$\pm$0.3 \\
r-STSF        & 5.09$\pm$0.15 & 1.04$\pm$0.28 & 52.8$\pm$0.9 \\
catch22       & 4.85$\pm$1.59 & 1.24$\pm$1.15 & 65.5$\pm$7.3 \\
ResNet        & 6.76$\pm$0.00 & 3.78$\pm$0.00 & 34.0$\pm$0.0 \\
\bottomrule
\end{tabular}
\end{center}
\vspace{-7mm}
\end{table}

\subsection{Sensor Importance}
\vspace{-1mm}
Table~\ref{tab:sensors} reports sensor-ablation results using QUANT. The accelerometer was the strongest valid single sensor, reaching 5.30\% EER. Its advantage over linear acceleration indicates that gravity and device-orientation cues contribute to the discriminative signal. The processed magnetometer was least informative. Raw magnetometer results are excluded because preliminary experiments showed context leakage from heading-related bias.

\begin{table}[ht!]
\def\arraystretch{1.0}
\setlength{\tabcolsep}{0.9em}
\begin{center}
\caption{Sensor-ablation results using QUANT with the double-window representation and all three axes. EER is computed by threshold sweeping; FRR and FAR are reported at the training-calibrated threshold (target FRR = 1\%). Values are mean $\pm$ standard deviation over 5 repeats.}
\label{tab:sensors}
\small
\begin{tabular}{lccc}
\toprule
Sensor & EER (\%) & FRR (\%) & FAR (\%) \\
\midrule
Accelerometer     & 5.30$\pm$0.28 & 1.40$\pm$0.46 & 57.3$\pm$3.3 \\
Linear accel.     & 8.95$\pm$0.70 & 1.16$\pm$0.48 & 74.8$\pm$1.5 \\
Gyroscope         & 9.60$\pm$0.76 & 1.22$\pm$0.48 & 77.3$\pm$1.5 \\
Proc. magnet.     & 16.3$\pm$0.37 & 1.49$\pm$0.63 & 81.5$\pm$2.7 \\
\bottomrule
\end{tabular}
\end{center}
\vspace{-7mm}
\end{table}

\section{\textcolor{black}{Discussion}}

\textcolor{black}{The results show that passive motion traces recorded during selfie capture contain measurable spoof-related and identity-related information. Motion is most useful when an attack disturbs expected handheld capture dynamics: stationary replay scenarios were rejected at FAR = 0\% by several methods, and QUANT+3-NN reduced FAR to 18.2\% on handheld replay scenarios. Temporally shifted sequences remained difficult, with the best FAR at 67.8\%, because they contain genuine handheld motion and differ mainly through timestamp misalignment.} \textcolor{black}{This confirms that motion alone cannot detect all injection-style attacks, especially real-device attacks that preserve authentic IMU traces, but can reject a measurable part of the attack space without adding user burden.} \textcolor{black}{The verification experiments show that the same traces carry user-related information in the same-device, same-session setting. WEASEL+MUSE reached 1.07\% EER on 9 channels, while the 3-axis accelerometer alone captured much of the useful signal. Raw acceleration outperformed linear acceleration, indicating that gravity and orientation cues matter beyond dynamic hand motion, and the double-window representation improved TSC-based verification by adding initial handling information after camera opening. These observations are relevant because accelerometers are widely available on budget devices, whereas gyroscopes and magnetometers may be absent or less reliable.}


\textcolor{black}{A key methodological result is that verification utility depends not only on closed-set classification accuracy but also on score calibration. EER reflects separability under threshold sweeping, whereas FRR and FAR in Tables~\ref{tab:tsc_best}--\ref{tab:sensors} are reported at a threshold calibrated for target FRR = 1\%. Thus, low EER with high calibrated FAR, as in QUANT, indicates separable but poorly graduated scores near low-FRR thresholds. WEASEL+MUSE kept these values closer, while ridge-based MR-HYDRA collapsed to FAR = 100\% despite moderate EER. Classification accuracy and threshold-based verification utility must therefore be evaluated separately.}

\textcolor{black}{The main limitations concern external validity. CanSelfie contains 30 participants, one smartphone model, one posture, and one session; cross-device, cross-session, posture, and environmental variation may affect motion patterns~\cite{stisen2015smart}. The attack-proxy set is small, especially for stationary and handheld replay, so percentages should be read as initial evidence rather than final attack-detection rates. The study does not exercise a real software-level injection pipeline, fuse motion scores with a face matcher, or address attackers who manipulate motion sensors jointly with the camera stream.}\textcolor{black}{ The absence of facial videos is also a privacy and consent-driven limitation: it supports motion benchmarking, but not direct video-motion fusion studies.}

\section{\textcolor{black}{Conclusion}}

\textcolor{black}{This paper introduced passive selfie-capture motion analysis as a biometric security task for mobile RIdV, motivated by the need for auxiliary evidence channels independent of the facial video stream and aligned with the defense-in-depth direction of ETSI TS 119 461~\cite{etsi119461} and CEN/TS 18099~\cite{cents18099}. The contributions are CanSelfie, a publicly available dataset of 375 multi-sensor selfie-capture sequences from 30 participants in a commercial RIdV workflow, stationary, handheld, and temporally shifted attack-proxy scenarios, and a benchmark of seven TSC and eight whole-series AD methods.} \textcolor{black}{The main findings are that short motion traces contain spoof-related information when attacks disturb handheld capture dynamics, carry user-related information in the same-device and same-session setting, and require score calibration when TSC models are adapted to biometric verification. Taken together, passive selfie-capture motion analysis is best understood as a low-friction auxiliary signal for layered RIdV decision-making, not as a standalone identity verifier or complete injection-attack detector. Future work should extend CanSelfie across devices, sessions, postures, and capture conditions, evaluate real video-injection and replay pipelines under a CEN/TS 18099-style protocol, and quantify fusion with facial recognition and presentation-attack detection.}

\section*{\textcolor{black}{Acknowledgment}}
\vspace{-2mm}
\textcolor{black}{This work was supported by the Research Council of Finland through the 6G Flagship Programme (Grant 369116), Business Finland through the HBIAS project (3092/31/2025), and the Technology Industries of Finland Centennial Foundation through a personal grant to Erkka Rantahalvari.}

{\small
\bibliographystyle{ieee}
\bibliography{references}
}

\end{document}